\documentclass[aps,prd,amsmath,amssymb,showpacs]{revtex4}

\usepackage{epsfig,float}
\usepackage{graphicx}
\usepackage{dcolumn}
\usepackage{morefloats}
\usepackage{color}
\usepackage{slashed}
\usepackage{bbm}

\begin{document}

\title{Some insights into the newly observed $Z_c(4100)$ in $B^0\to \eta_c K^+ \pi^-$ by LHCb }

\author{Qiang Zhao$^{1,2}$\footnote{{\it Email address:} zhaoq@ihep.ac.cn} }

\affiliation{$^1$ Institute of High Energy Physics and Theoretical Physics Center for Science Facilities,
        Chinese Academy of Sciences, Beijing 100049, China}

\affiliation{$^2$  School of Physical Sciences, University of Chinese Academy of Sciences, Beijing 100049, China}


\begin{abstract}
The newly reported exotic signal $Z_c(4100)$ by the LHCb Collaboration in the invariant mass spectrum of $\eta_c\pi^-$ in $B^0\to \eta_c K^+ \pi^-$ has been a new experimental evidence for an exotic meson containing four constituent quarks. Although the present experimental information is very limited, we show that its correlations with some existing exotic candidates can be recognized. This signal can be either caused by final state interaction effects or a $P$-wave resonance state arising from the $D^*\bar{D}^*$ interaction. For the latter option its neutral partner will have exotic quantum numbers of $I^G(J^{PC})=1^-(1^{-+})$. This signal, if confirmed, would provide important clues for dynamics for producing multiquark systems in $B$ meson decays and $e^+e^-$ annihilations.

\end{abstract}
\date{\today}

\maketitle
\section{Background}

During the past decade there have been a sizeable number of new hadron states observed in experiment. Some of these states have been ideal candidates for QCD exotic hadrons and they have motivated a lot of theoretical and further experimental studies of their internal structures. In particular, the observations of a series of charged heavy quarkonium states~\cite{Choi:2007wga,Mizuk:2008me,Belle:2011aa,Ablikim:2013mio,Liu:2013dau,Xiao:2013iha,Ablikim:2013wzq,Ablikim:2013xfr,Ablikim:2013emm,Aaij:2014jqa,Lees:2011ik}
have initiated tremendous efforts on decoding their structures and underlying dynamics (see several recent reviews on the relevant theoretical and experimental progresses~\cite{Chen:2016qju,Chen:2016spr,Guo:2017jvc,Lebed:2016hpi,Esposito:2016noz,Olsen:2017bmm}).

The decay process $B^0\to \eta_c K^+ \pi^-$ was recently published by the LHCb Collaboration as the first measurement~\cite{Aaij:2018bla}. An enhancement with a mass of $4096\pm{20}^{+18}_{-22}$ MeV and a width of $152\pm 58^{+60}_{-35}$ MeV is found in the invariant mass spectrum of $\eta_c\pi^-$. The significance of this signal is about three standard deviations which is not strong enough for a conclusion, but is interesting for exploring possible implications arising from such a phenomenon. The advantage of heavy hadron weak decays is that it can access various quantum numbers and multiquark combinations in the final state. In the case of $B^0\to \eta_c K^+ \pi^-$ the main quasi-two-body decay actually is $B^0\to \eta_c K^{*0}$ as shown by the detailed analysis~\cite{Aaij:2018bla}. Here, $K^{*0}$ denotes excited kaon resonances that can decay into $K^+\pi^-$. In contrast, any combination of $\eta_c K^+$ or $\eta_c\pi^-$ would signal evidences for exotic mesons. It should be noted that evidences for exotic candidates $Z_c(4430)$ and $Z_c(4200)$ have been reported by the Belle Collaboration in $B^0\to J/\psi K^+\pi^-$ in the invariant mass spectrum of $J/\psi\pi^-$~\cite{Chilikin:2014bkk}. The $Z_c(4100)$ enhancement observed in the invariant mass spectrum of $\eta_c\pi^-$, if confirmed, would then provide very useful information for the multiquark dynamics.

At this moment there are questions raised by this possible exotic candidate. Firstly, why this state has a mass far higher than the thresholds of $D\bar{D}$ ($\sim 3740$ MeV)? Is it the ground state? Or is it a resonance state? If it is a resonance state, why the ground state, which is unlikely to be lower than the $\eta_c\pi$ threshold, has not been seen in the $\eta_c\pi$ invariant mass spectrum? Meanwhile, if indeed the $Z_c(4100)$ as an exotic state containing four quarks does exist, one would have to consider its relation with the better established $Z_c(3900)$ and $Z_c(4020)$. In particular, their productions in $B$ meson decays and their relations with $Z_c(4100)$ and other reported charged charmonium-like states should be investigated. In principle, these challenging questions should be addressed by any proposal for its nature. In Ref.~\cite{Voloshin:2018vym} it is proposed that $Z_c(4100)$ and $Z_c(4200)$ can be interpreted as hadro-quarkonium states. Further studies to either confirm or deny this signal in other processes will advance our understanding of the structures and production mechanisms for charmonium-like states in $B$ meson decays.

In this work we try to address several crucial issues relevant to the understanding of the $Z_c(4100)$ signal. This will be useful for future experimental investigations. As follows, we first note several crucial issues that one should be aware of in Sec. II.  We then propose an explanation for the $Z_c(4100)$ structure and discuss criteria for further experimental studies in Sec. III. A brief summary is presented in Sec. IV.

\section{Puzzles and questions}

\subsubsection{Quantum numbers and possible decay channels}

\begin{figure}[h]
\begin{center}
\includegraphics[scale=0.3]{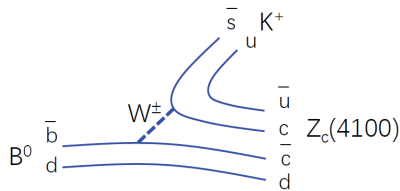}
\caption{ Schematic diagram for the production of $Z_c(4100)$ as a four-quark state containing $c\bar{c}d\bar{u}$ in $B^0\to \eta_c K^+ \pi^-$. }
\label{fig-1}
\end{center}
\end{figure}

The production of $Z_c(4100)$ in $B^0\to \eta_c K^+ \pi^-$  is referred to the illustrative process of Fig.~\ref{fig-1} as the leading production mechanism. Notice that the final state $\bar{c}d$ is initially in a color singlet. It means that the multi-quark state $c\bar{c}d\bar{u}$ is produced by the strong interaction between two color singlets $\bar{c}d$ and $c\bar{u}$. The quantum numbers accessible for $\eta_c\pi^-$ can be $I^G(J^P)=1^-(0^+)$  with an $S$-wave interaction or $1^-(1^-)$ with a $P$-wave interaction. For this four-quark system it implies that the $S$-wave decay channels of $Z_c(4100)^-\to J/\psi\rho^-$, $D^0D^-$ and $D^{*0}D^{*-}$ are allowed for $J^P=0^+$. One also notices that the $P$-wave decay into $\chi_{c1}\pi^-$ is also allowed. For $J^P=1^-$, its decays into $\chi_{c1}\pi^-$ and $J/\psi a_0^-$ are via the $S$ wave, while the decay channels of $J/\psi\pi^-$, $J/\psi\rho^-$, $D^0D^-$, $D^{*0}D^{*-}$, $D^0\bar{D}^{*-}$ and $D^{*0}\bar{D}^-$ are via the $P$ wave. In Table~\ref{tab-1} we list the kinematically accessible decay channels for the charged $Z_c(4100)$ with these two most likely quantum numbers. We will discuss the neutral channels later.

\begin{table}[htbp]
  \centering
  \caption{ Possible quantum numbers and corresponding decay channels for the charged exotic candidate $Z_c(4100)$. To save space, the open charm decays of the positively charged $Z_c(4100)^+$ are implicated. The abbreviations ``HC" and ``OC" denote hidden charm and open charm decays, respectively, while the short dash ``-" means the corresponding decay channels are either non-existing or forbidden by the kinematics. }
    \begin{tabular}{c|c|c|c|c}
    \hline
     $I^G(J^P)$  &  \multicolumn{2}{c|}{$1^-(0^+)$}
&    \multicolumn{2}{c}{$1^-(1^-)$}
  \\
     \hline
     & HC  & OC  & HC & OC \\
     \hline
    $S$ wave & $\eta_c\pi^\pm$, $J/\psi\rho^\pm$ & $D^0D^-,  \ D^{*0}D^{*-}$ & $\chi_{c1}\pi^\pm$, $J/\psi a_0^\pm$ & - \\
    $P$ wave & $\chi_{c1}\pi^\pm$, $J/\psi a_0^\pm$ &  - & $\eta_c\pi^\pm$, $J/\psi\pi^-$, $J/\psi\rho^\pm$ & $ D^0D^-,  \ D^{*0}D^{*-},  \ D^0D^{*-}$ \\
    \hline
\end{tabular}%
  \label{tab-1}%
\end{table}

For $J^P=0^+$, the decay channels of pseudoscalar meson pairs (noted as ``$PP$") or vector meson pairs (noted as ``VV") can be related to each other by the heavy quark spin symmetry (HQSS) which leads to $BR(\eta_c\pi^-)/BR(J/\psi\rho)\simeq q(\eta_c\pi)/3 q(J/\psi\rho)\simeq 0.48$. Also, the following branching ratio fractions can be expected: $BR(\eta_c\pi^-)/BR(D^0D^-)\simeq q(\eta_c\pi)/q(D^0D^-)\simeq 1.14$ and $BR(J/\psi\rho)/BR(D^*\bar{D}^*)\simeq q(J/\psi\rho)/q(D^*\bar{D}^*)\simeq 1.44$, where $q$ denotes the three-vector momenta for the final state meson in each channel in the $Z_c(4100)$ rest frame. These ratios suggest that the $J/\psi\rho$ channel should also be an important decay channel for the $Z_c(4100)$ with $J^P=0^+$. This channel would be useful for further clarifying the property of this signal.

Note that for the $J^P=0^+$ option, $Z_c(4100)$ can decay into $\chi_{c1}\pi$ via the $P$ wave. In case that the HQSS is still reasonably respected, one may expect that $BR(J/\psi\rho)/BR(\chi_{c1}\pi)\simeq q(J/\psi\rho)/q(\chi_{c1}\pi)^3\simeq 2.8$. It suggests that the signal for $Z_c(4100)$ in $B^0\to \chi_{c1} K^+ \pi^-$ may provide a lower limit for the signal significance to be observed in experiment. In Ref.~\cite{Mizuk:2008me} two charged charmonium-like states were reported in the invariant mass spectrum of $\chi_{c1}\pi^+$ in $\bar{B}^0\to K^-\pi^+\chi_{c1}$, i.e. $Z_c(4051)$ and $Z_c(4248)$. Their masses and widths are $m(Z_c(4051))=(4051\pm 14^{+20}_{-41})$ MeV, $\Gamma(Z_c(4051))=(82^{+21 \ +17}_{-47 \ -22})$ MeV,
$m(Z_c(4248))=(4248^{+44 \ +180}_{-29 \ -35})$ MeV, and $\Gamma(Z_c(4248))=(177^{+54 \ +39}_{-316 \ -61})$ MeV. As shown by the analysis of Ref.~\cite{Mizuk:2008me} their total spin can be either 0 or 1 which result in insignificant differences in the fitting. This makes it quite possible that $Z_c(4100)$ and $Z_c(4051)$ are the same state. Although BaBar did not confirm these two states in their analysis~\cite{Lees:2011ik}, the relatively low statistics in both analyses make it desirable to have improved analysis of this channel at LHCb. Also note that if the quantum numbers of $J^P=1^-$ are favored for $Z_c(4100)$, the qualitative branching ratio fraction between the $J/\psi\rho$ and $\chi_{c1}\pi$ channel will become $BR(J/\psi\rho)/BR(\chi_{c1}\pi)\simeq q(J/\psi\rho)^3/q(\chi_{c1}\pi)\simeq 0.48$ because of the change of the partial waves in its decays as noted in Table~\ref{tab-1}. Thus, a comparison between the $J/\psi\rho$ and $\chi_{c1}\pi$ channels will be useful for establishing the quantum numbers of $Z_c(4100)$ if it does exist.

Furthermore, if the $Z_c(4100)$ does exist, one would have to consider its relation with the better established $Z_c(3900)$ and $Z_c(4020)$ in both $B$ meson decays and $e^+e^-$ annihilations. Note that the branching ratio $BR(B^0\to \eta_c K^+\pi^-)=(5.73\pm 0.24\pm 0.13\pm 0.66)\times 10^{-4}$ is extracted by normalizing it to that for $B^0\to J/\psi K^+\pi^-$, $BR(B^0\to J/\psi K^+\pi^-)=(1.15\pm 0.05)\times 10^{-3}$~\cite{Tanabashi:2018oca}. Although there are still large uncertainties for unknown contributions to $B^0\to \eta_c K^+\pi^-$ which is estimated by the last error, it shows that the branching ratio for $B^0\to \eta_c K^+\pi^-$ turns out to be smaller than that for $B^0\to J/\psi K^+\pi^-$. Then, the question is, Should one expect $Z_c(3900)$ and/or $Z_c(4020)$ to appear in the invariant mass spectrum of $J/\psi\pi^-$ in $B^0\to J/\psi K^+\pi^-$? In principle, the production mechanism for $Z_c(4100)$ in the $B^0$ decays should also work for $Z_c(3900)$ and $Z_c(4020)$. In the recent analysis by the D0 Collaboration~\cite{Abazov:2018cyu} the channel $\bar{B}^0\to J/\psi K^-\pi^+$ is investigated and no sign for the $Z_c(3900)$ appears in the invariant mass spectrum of $J/\psi\pi^+$. Although the data show some structure around $4.0\sim 4.05$ GeV and $4.15\sim 4.20$ GeV, the low statistics do not allow any conclusion on their properties at all. However, it is interesting that the signals for $Z_c(3900)$ appear in semi-inclusive weak decays of b-flavored hadrons and strongly correlated with the production of $Y(4260)$ as shown by the analysis of Ref.~\cite{Abazov:2018cyu}. Actually, relatively higher statistics analysis of $\bar{B}^0\to J/\psi K^-\pi^+$ was given by the Belle Collaboration~\cite{Chilikin:2014bkk}. It was found in Ref.~\cite{Chilikin:2014bkk} that two charged charmonium states, $Z_c(4200)$ and $Z_c(4430)$, instead of $Z_c(3900)$ and $Z_c(4020)$, are needed in the invariant mass spectrum of $J/\psi\pi^+$ in the amplitude analysis. It seems that there is also no need for the $Z_c(4100)$ to be present in the $J/\psi\pi^+$ spectrum based on the present statistics. While such an observation may disfavor $Z_c(4100)$ to have $J^P=1^-$, it is also possible that the present statistics are not sufficient for singling out its signal.

Keeping the above observations and questions in mind, we try to propose a possible explanation for the observed $Z_c(4100)$.

\section{A possible explanation}

It would be useful to take the relatively well-established $Z_c(3900)$ and $Z_c(4020)$ as a reference for understanding the $Z_c(4100)$ enhancement. In particular, the mass of $Z_c(4100)$ is relatively close to the threshold of $D^*\bar{D}^*$ in either an $S$ or $P$ wave. If we argue that so far not so many multiquark states have been observed in experiment, we then try to seek a rather normal interpretation for a signal like this.

Noticing that the mass of $Z_c(4100)$ is not far away from the $D^*\bar{D}^*$ threshold, it would be possible that the $D^*\bar{D}^*$ final state interactions can produce structures that may be visible in experiment. Such a mechanism is a rather common phenomenon in hadron-hadron interactions. With a strong attractive near-threshold $S$-wave interaction, a hadronic molecule state can even be dynamically generated, and experimental candidates include some of these recently observed so-called ``$XYZ$" states. An up-to-date review of hadronic molecules can be found in Ref.~\cite{Guo:2017jvc}.

\begin{figure}[h]
\begin{center}
\includegraphics[scale=0.3]{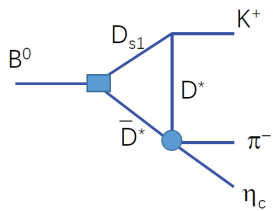}
\caption{The $Z_c(4100)$ enhancement produced by the $P$-wave $D^*\bar{D}^*$ interactions in $B^0\to \eta_c K^+ \pi^-$ via the intermediate $D_{s1}(2460)\bar{D}^*$ ($D_{s1}(2536)\bar{D}^*$) channel.  }
\label{fig-2}
\end{center}
\end{figure}

In Fig.~\ref{fig-2} a schematic illustration is plotted. The $B^0$ can decay into $\eta_c K^+\pi^-$ via intermediate $D_s^{(*)}\bar{D}^*$ with the rescattering of $D^*\bar{D}^*$ into $\eta_c\pi^-$ to cause the $Z_c(4100)$ structure. The measured branching ratios for $B^0\to D_s^{(*)}\bar{D}^*$ are dominant ones at the order of $10^{-3}\sim 10^{-2}$~\cite{Tanabashi:2018oca}. In particular, the branching ratio $BR(B^0\to D_s^{+}(2457)D^{*-})=(9.3\pm 2.2)\times 10^{-3}$~\cite{Tanabashi:2018oca} is much larger than $BR(B^0\to \eta_c K^+ \pi^-)=(5.73\pm 0.24\pm 0.13\pm 0.66)\times 10^{-4}$~\cite{Aaij:2018bla}.

We consider possibilities arising from such a rescattering phenomenon for these two possible quantum numbers, i.e. $J^P=0^+$ or $1^-$.

\subsection{$J^P=0^+$}

Taking the $Z_c(4020)$ as a reference, the $D^*\bar{D}^*$ interactions can be strongly repulsive if they couple into total spin 0 state~\footnote{The total spin 0 in a relative $S$ wave for the $D^*$ and $\bar{D}^*$ suggests that if the light $q\bar{q}$ couple to spin 0, the total spin between the heavy $c$ and $\bar{c}$ will also couple to 0.}. This can be understood by the spin-flavor interactions between $D^*\bar{D}^*$ via pion exchange. The typical pion exchange potential involving the light quarks can be expressed as:
\begin{equation}
    \hat{H}_\pi=g \vec\sigma_1\cdot\vec\sigma_2 \vec\tau_1\cdot\vec\tau_2 \ ,
\end{equation}
where $\vec\sigma_1$ and $\vec\sigma_2$ ($\vec\tau_1$ and $\vec\tau_2$) are the spin (isospin) operators at the $D^* D^*\pi$ and $\bar{D}^*\bar{D}^*\pi$ vertices, respectively, which involves the light quark degrees of freedom. For the $Z_c(4020)$ as an $S$-wave $D^*\bar{D}^*$ hadronic molecule, the potential sign and relative strength is determined by
\begin{equation}
    \frac{\langle \hat{H}_\pi(Z_c(4100))\rangle}{\langle \hat{H}_\pi(Z_c(4020))\rangle}\simeq \frac{1}{-3} \ ,
\end{equation}
which means that the attractive potential for $Z_c(4020)$ with $J^P=1^+$ would become repulsive for $Z_c(4100)$ with $J^P=0^+$ ($S$ wave) and $1^-$ ($P$ wave) with total spin 0 (For the case of $P$ wave with total spin 1 we will discuss it in the next subsection). The brackets mean to take the mean value of the pion exchange potential between the wavefunctions for the initial and final $D^*$ and $\bar{D}^*$ states. Note that the interaction between the $D^*\bar{D}^*$ has the same order of magnitude as that for $Z_c(4020)$. The consequence is that such a repulsive interaction can possibly produce observable effects in its rescatterings into $\eta_c\pi^-$, $D\bar{D}$, $J/\psi\rho$, and even $D^*\bar{D}^*$. If this scenario follows, it suggests that the $Z_c(4100)$ enhancement is unlikely to be a genuine state, but rather likely to be a final state interaction effect.

The production of $Z_c(4100)$ in different processes can be prospected. In the hadronic molecular picture, the $S$-wave $D\bar{D}$ system is different from either $D\bar{D}^*+c.c.$ or $D^*\bar{D}^*$ due to the absence of the long-range pion exchange potential between the $D$ and $\bar{D}$ meson pair. However, a tetraquark state which can strongly couple to $D\bar{D}$ in an $S$ wave is anticipated in the tetraquark model. In order to distinguish the final state rescattering effect from the tetraquark state, a measurement of the $D^- D^0$ invariant mass spectrum in $B^0\to K^+ D^- D^0$ would be useful. The absence of the state with $J^P=0^+$ near the $D\bar{D}$ threshold, but the presence of an enhancement near 4.1 GeV would be an evidence for the rescattering effect for $D^*\bar{D}^*\to D\bar{D}$, instead of a genuine tetraquark state. Note that this channel has been measured by BaBar~\cite{delAmoSanchez:2010pg} with a branching ratio $BR(B^0\to B^0\to K^+ D^- D^0)=(2.7\pm 1.1)\times 10^{-4}$~\cite{Tanabashi:2018oca}. This branching ratio is compatible with $BR(B^0\to \eta_c K^+ \pi^-)=(5.73\pm 0.24\pm 0.13\pm 0.66)\times 10^{-4}$.

With the significant branching ratio $BR(B^0\to D_s^{+}(2457)D^{*-})=(9.3\pm 2.2)\times 10^{-3}$\cite{Tanabashi:2018oca}, it can be expected that the intermediate $D^{*0}D^{*-}$ rescatterings into $D\bar{D}$ and $\eta_c\pi^-$ final states can contribute to the branch ratios for these two channels. Although the detailed quantitative results need evaluations of the loop contributions, one can still gain a qualitative idea about their relative strength:
\begin{equation}
R\equiv \frac{T(D^*\bar{D}^*\to D\bar{D})}{T(D^*\bar{D}^*\to \eta_c\pi)}\simeq \frac{g_{D^*D\pi}^2}{g_{D^*D\pi}g_{\eta_c D^*\bar{D}}} \ ,
\end{equation}
where $g_{D^*D\pi}\simeq 5.96$ and $g_{\eta_c D^*\bar{D}}=2g_2(m_{\eta_c}m_D m_{D^*})^{1/2}\simeq 7.68$ with $g_2\equiv \sqrt{m_{J/\psi}}/2m_D f_{J/\psi}$ denote the coupling constants that can be either determined by the experimental data or relations established based on the heavy quark spin symmetry~\cite{Casalbuoni:1996pg}. At the threshold of $D^*\bar{D}^*$, the ratio $R\simeq 0.78$ can be extracted. Note that the $\eta_c\pi$ channel will experience suppressions from the $D$ meson propagator. It makes the $D\bar{D}$ interesting for the search for further evidences for $Z_c(4100)$.

\subsection{$J^P=1^-$}

Supposing the $Z_c(4100)$ originates from the $P$-wave $D^*\bar{D}^*$ interaction with $I=1$, the most interesting implication of this possible assignment of quantum numbers is that its neutral partner can have $J^{PC}=1^{-\pm}$, in which the quantum number $J^{PC}=1^{-+}$ is for an exotic state beyond the $q\bar{q}$ scenario. Whether the neutral $Z_c(4100)^0$ can decay into $\eta_c\pi^0$ or not will determine its quantum number to be either $J^{PC}=1^{--}$ or $1^{-+}$. This can be regarded as an additional constraint on the nature of $Z_c(4100)$ enhancement. Therefore, a search for the $Z_c(4100)$ signal in the neutral decay channel, either $\eta_c\pi^0$ or $J/\psi\rho^0$ will be helpful for clarifying its nature.

A spin decomposition of the neutral $D^*\bar{D}^*$ in a $P$-wave can explicitly show the relative coupling strengths to the hidden charm decay channels:
\begin{eqnarray}
|D^*\bar{D}^*\rangle^{S=0}_{1^{--}}&=&\frac{\sqrt{3}}{2}|0\otimes 1\rangle -\frac{1}{6}|1\otimes 0\rangle +\frac{1}{2\sqrt{3}}|1\otimes 1\rangle -\frac{\sqrt{5}}{6}|1\otimes 2\rangle \ , \label{S0}\\
|D^*\bar{D}^*\rangle^{S=2}_{1^{--}}&=&\frac{\sqrt{5}}{3}|1\otimes 0\rangle +\frac{1}{2}\sqrt{\frac{5}{3}}|1\otimes 1\rangle +\frac{1}{6}|1\otimes 2\rangle \ ,\label{S2}\\
|D^*\bar{D}^*\rangle^{S=1}_{1^{-+}}&=&\frac{1}{\sqrt{2}}|0\otimes 1\rangle + \frac{1}{\sqrt{2}}|1\otimes 1\rangle \ . \label{S1}
\end{eqnarray}
Similar spin-decomposition studies can be found in Ref.~\cite{Du:2016qcr}, and we follow the convention defined in Ref.~\cite{Du:2016qcr} to note the spin of the final state heavy charmonium and the total angular momentum carried by the final-state light quarks as $|s_Q\otimes s_l\rangle$, with $\vec{s}_l=\vec{s}_q+\vec{l}$. Noticing that the configuration of $|D^*\bar{D}^*\rangle^{S=2}_{1^{--}}$ (Eq.~(\ref{S2})) does not contain $\eta_c\pi$ in the decomposition, it can be ruled out from the present observation.

For the $S$-wave coupling it has been recognized that the allowed quantum numbers for the neutral $D^*\bar{D}^*$ system can be $J^{PC}=0^{++}$, $1^{+-}$, and $2^{++}$. Given the long-range pion exchange potential, whether this system has an attractive or repulsive interaction would rely on their isospin quantum number in the flavor-spin coupling scenario. At this moment the only candidate for the $S$-wave $D^*\bar{D}^*$ threshold state is $Z_c(4020)$ observed in $e^+e^-\to h_c\pi\pi$ and $D^*\bar{D}^*\pi$ at BESIII, which belongs to the $I^G (J^{PC})=1^+ (1^{+-})$ category. By assuming that the pion exchange potential provides an attractive interaction between $D^*\bar{D}^*$ to form $Z_c(4020)$, we deduce that the attractive potential has the sign determined by $\langle\vec\sigma_1\cdot\vec\sigma_2\rangle\langle \vec\tau_1\cdot\vec\tau_2\rangle=(+1)(+1)=1$~\footnote{In Ref.~\cite{Du:2016qcr} the authors show that the state of $I^G (J^{PC})=1^+ (1^{+-})$ for the $D^*\bar{D}^*$ system is more likely to be bound than other configurations based on the inclusion of more sophisticated meson exchange potentials in addition to the pion exchange.}. This may lead to the formation of the resonance state with $I=1$ and $J^{PC}=1^{-+}$ made of the $P$-wave neutral $D^*\bar{D}^*$. It also implies that the configuration of Eq.~(\ref{S0}) but with $I=0$ could have attractive interactions, $\langle\vec\sigma_1\cdot\vec\sigma_2\rangle\langle \vec\tau_1\cdot\vec\tau_2\rangle=(-3)(-3)=+9$, to form a molecule-like state with $J^{PC}=1^{--}$. This turns out to be a rather strong attraction and the state has the same quantum numbers as a vector charmonium state. This is consistent with the recent study of Ref.~\cite{Du:2016qcr}, where
the coupled-channel calculation indicates that the $\psi(4040)$, which is assigned as a $(3^1S_1)$ state in the quark model, has a strong coupling to $D^*\bar{D}^*$. Thus, its wavefunction can be affected strongly by the renormalization of the $P$-wave $D^*\bar{D}^*$ interaction.

\section{Summary and propects}

As a brief summary for the possible assignments for $Z_c(4100)$, we propose that it could be a rescattering effect arising from an $S$-wave $D^*\bar{D}^*$ rescattering with $I^{(G)}(J^{P(C)})=1^{(-)}(0^{+(+)})$, which does not signal a genuine state. It is also possible that the structure corresponds to a resonance produced by the $P$-wave $D^*\bar{D}^*$ interaction of which the neutral partner has exotic quantum number $I^G(J^{PC})=1^-(1^{-+})$. This can be regarded as an excited resonance state of $Z_c(4020)$. In contrast, the quantum number of $I^G(J^{PC})=1^+(1^{--})$ is forbidden by the $C$ parity.

In order to further clarify the nature of $Z_c(4100)$ and search for evidences for other possible $Z_c$ states, we suggest a combined analysis of $e^+e^-\to D\bar{D}\pi$, $D\bar{D}\pi\pi$, and $D\bar{D}\pi\pi\pi$ at the resonance peaks, e.g. $E_{c.m.}=4.43$ and 4.66 GeV, can have direct access to $I=1$ charmonium-like states with $J^P=0^+, \ 1^-, \ 1^+, \ 2^+$, etc. For the hidden charm decay channels, it is unlikely that a sizeable $I=1$ amplitude can be directly produced by the electromagnetic (EM) current in the charmonium energy region, not mentioning that in such a case the $c\bar{c}$ created from the vacuum will be highly suppressed. Therefore, the production of the $I=1$ charmonium-like states will always be associated by isovector states such as $\pi$ or $\rho$ mesons. With such a consideration, a combined analysis of the reaction channels, such as $e^+e^-\to J/\psi \pi\pi$, $J/\psi\pi\pi\pi$, $\eta_c\pi\pi$, $\eta_c\pi\pi\pi$, etc., should be very much useful. Meanwhile, the open charm reaction channels $e^+e^-\to \rho D\bar{D}$, $\rho D\bar{D}^*+c.c.$, and $\rho D^*\bar{D}^*$, can serve as probing channels for the $Z_c$ states with the quantum numbers implied by the LHCb data. For the exotic states of $J^{PC}=1^{-+}$ a proposal for studying its nature in $e^+e^-$ annihilations can be found in Ref.~\cite{Wang:2014wga}, where the radiative transition channels of vector charmonia were proposed for further investigations. We anticipate that future data and analyses from LHCb, BESIII, and Belle-II can provide deeper insights into the nature of these charged charmonium-like states.

\section*{Acknowledgment}

This work is supported, in part, by the National Natural Science Foundation of China (Grant Nos. 11425525 and 11521505), DFG and NSFC funds to the Sino-German CRC 110 ``Symmetries and the Emergence of Structure in QCD'' (NSFC Grant No. 11261130311), National Key Basic Research Program of China under Contract No. 2015CB856700.

\end{document}